\documentclass[aps,prd,twocolumn,showpacs,preprintnumbers,nofootinbib,amsmath,amssymb]{revtex4}

\usepackage{graphicx}
\usepackage{dcolumn}
\usepackage{bm}
\usepackage{amsmath}

\newcommand{\beq}{\begin{eqnarray}}
\newcommand{\eeq}{\end{eqnarray}}

\newcommand{\real}{{\sf I}\kern-.12em{\sf R}}
\newcommand{\comp}{{\sf I}\kern-.50em{\sf C}}
\newcommand{\unity}{{\sf I}\kern-.54em{\sf 1}}

\def\spose#1{\hbox to 0pt{#1\hss}}
\def\ltapprox{\mathrel{\spose{\lower 3pt\hbox{$\mathchar"218$}}
 \raise 2.0pt\hbox{$\mathchar"13C$}}}

\begin{document}

\title{Chiral properties of strong interactions in a magnetic background}
\author{Massimo D'Elia and Francesco Negro}
\affiliation{Dipartimento di Fisica, Universit\`a
di Genova, Via Dodecaneso 33, 16146 Genova, Italy\\
INFN, Sezione di Genova, Via Dodecaneso 33, 16146 Genova, Italy\\
}
\date{\today}

\begin{abstract}
We investigate the chiral properties of QCD in presence of a magnetic background
field and in the low temperature regime, by lattice numerical simulations of $N_f = 2$ QCD.
We adopt a standard staggered discretization, with a pion mass around 200 MeV, and explore
a range of magnetic fields  $(180\ {\rm MeV})^2 \leq |e|B \leq (700\ {\rm MeV})^2$, in which
we study magnetic catalysis, i.e. the increase of chiral symmetry breaking induced
by the background field.
We determine the dependence of the chiral condensate on the external field, 
compare our results with existing model predictions and show that a substantial
contribution to magnetic catalysis comes from the modified distribution of non-Abelian
gauge fields, induced by the magnetic field via dynamical quark loop effects.
\end{abstract}

\maketitle

\section{Introduction}

The study of strong interactions in presence of a strong background magnetic
field has attracted increasing attention in the recent past. On one side
the issue is of great phenomenological relevance: magnetic 
fields of the order of $10^{16}$ Tesla, i.e. $\sqrt{|e|B} \sim 1.5$ GeV) may have been 
produced at the cosmological electroweak phase transition~\cite{Vachaspati:1991nm} 
and they may have
influenced subsequent strong interaction dynamics, including the 
confinement/deconfinement transition. Slightly lower fields 
are expected to be produced in non-central heavy
ion collisions (up to $10^{14}$ Tesla at RHIC and up to $\sim  10^{15}$
Tesla at LHC~\cite{heavyionfield1,heavyionfield2}), where they may give 
rise to new phenomenology, the so-called "chiral magnetic effect", 
capable of revealing the presence of deconfined matter
and of non-trivial topological vacuum fluctuations~\cite{cme0,cme1,star}.
Finally, magnetic fields of the
order of $10^{10}$ Tesla are expected to be present 
in a class of neutron stars known as
magnetars~\cite{magnetars} (for a recent review see Ref.~\cite{mereghetti}).

On the other side, a background magnetic field (electro-magnetic or chromo-magnetic) 
may serve as yet another parameter
to probe the structure of the QCD vacuum and of the QCD phase diagram, on the same
footing with other interesting external 
conditions such as a baryon chemical potential.
Many studies~\cite{salam,linde,Kawati:1983aq,Klevansky:1989vi,Suganuma,Klimenko,Schramm,
Klimenko:1993ec,Gusynin:1994re,Gusynin:1994xp,Shushpanov:1997sf,Babansky:1997zh,Ebert,Goyal:1999ye,Agasian:1999sx,
Ebert:2001ba,Kabat:2002er,Miransky:2002rp,Cohen:2007bt,Johnson:2008vna,Rojas,Bergman:2008qv,Zayakin:2008cy,Evans:2010xs,mizher,Nam} 
have investigated the chiral properties of the 
theory and what is generally known as magnetic catalysis, 
consisting in an enhancement of chiral symmetry breaking and 
in spontaneous mass generation
induced by the magnetic field, a phenomenon predicted from different
low energy models and approximations of QCD and related 
to the dimensional reduction taking place in the dynamics of particles moving in a 
strong external magnetic field~\cite{Gusynin:1994re,Gusynin:1994xp,Miransky:2002rp}.
 More recently, the issue of the influence
of a magnetic field on the deconfinement transition 
has been investigated
by means of both lattice QCD simulations and low energy models of strong 
interactions~\cite{Klimenko,Nam,agasian,fraga1,Avancini,Ayala,NJL,fukushima,fraga2,demusa,gatto1,frolov,sadooghi,gatto2,
Preis:2010cq}: there is converging evidence that a magnetic field
leads to an increase of both the strength and the temperature of the transition;
in the case of a chromo-magnetic field, instead, numerical simulations show
a decrease of the transition temperature~\cite{bari1,bari2}. Finally, conjectures have been
proposed according to which a strong enough magnetic field may induce the 
appearance of new superconductive phases~\cite{supercond1,supercond2,supercond3}.

In the present study we address the issue of magnetic catalysis, 
presenting the first study of such phenomenon by lattice QCD simulations
which include the contribution of dynamical quarks; previous numerical
studies indeed have only considered the effect of the magnetic field
on quenched configurations~\cite{itep1,itep2}.
In particular, we have considered
QCD at zero or low temperature, with two dynamical flavors carrying different electric charges, 
corresponding respectively to the $u$ and $d$ quark charges, and coupled to a background
constant and uniform magnetic field. We have adopted a standard rooted staggered fermion
discretization,with a (Goldstone) pion mass of about 200 MeV.

One of the purposes of our investigation is to obtain information about the 
dependence of the chiral condensate on the magnetic field, and compare it 
with various existing model and low energy predictions.
A second purpose that we have is to understand which part of magnetic catalysis
is a purely tree level effect, due to the fact that quarks propagate in a modified background
obtained by adding the U(1) field to the non-Abelian gauge configurations, 
and which part is due to a modification of the non-Abelian fields themselves, 
induced by the loop effects of dynamical quarks coupled to the 
magnetic background. Both effects can in principle modify the 
spectrum of the Dirac operator, leading to a larger density of 
eigenvalues around zero, hence to an increase of the chiral condensate
via the Banks -- Casher relation~\cite{banks}.

To that aim, one could compare with existing investigations
of magnetic 
catalysis, based on SU(2) and SU(3) gauge configurations sampled by pure gauge simulations~\cite{itep1,itep2}:
however that would not be completely satisfactory and would also be 
difficult because of different scale settings and renormalization
effects. 
What we will do instead is to try separating the two different effects, by
inserting alternatively the magnetic field only in the
computation of the quark propagator (sampling in this case configurations without the presence of the magnetic
field), or only in the sampling measure, i.e. in the fermion determinant, without affecting the quark propagator computation.
We will call "valence" catalysis the first contribution and "dynamical" catalysis the second:
both of them will be compared with the full increase of the chiral condensate,
obtained when the magnetic field is inserted directly both in the fermion determinant and in the computation 
of the quark propagator.
As we will show, the purely dynamical contribution corresponds
to a considerable part of the total increase in the quark condensate.

The paper is organized as follows.  In Sec.~\ref{setup} we give some details
about our lattice discretization of QCD in presence of a magnetic field and about our
numerical setup. In Section~\ref{results} we present our numerical results and
finally, in Section~\ref{conclusions}, we give our conclusions.

\section{Numerical Setup}
\label{setup}

The discretization of $N_f = 2$ QCD in presence of a magnetic background field 
adopted in the present work is similar to that reported in Ref.~\cite{demusa}.
In particular, partition function of the (rooted) staggered fermion
discretized version of the theory in presence of a non-trivial
electro-magnetic (e.m.) background field and with  different electric charges for the two flavors, 
$q_u = 2|e|/3$ and $q_d = -|e|/3$ ($|e|$ being the elementary charge), is written
as:
\beq
Z(T,B) \equiv \int \mathcal{D}U e^{-S_{G}} 
\det M^{1\over 4} [B,q_u]
\det M^{1\over 4} [B,q_d]
\:
\label{partfun1}
\eeq
\begin{eqnarray}
M_{i,j} [B,q] &=& a m \delta_{i,j} 
+ {1 \over 2} \sum_{\nu=1}^{4}\eta_{i,\nu} \left(
\vphantom{ U^{\dag}_{i-\hat\nu,\nu} }
u(B,q)_{i,\nu} U_{i,\nu} \delta_{i,j-\hat\nu}
\right. \nonumber \\ && - \left.
u^*(B,q)_{i - \hat\nu,\nu} U^{\dag}_{i-\hat\nu,\nu}\delta_{i,j+\hat\nu} 
\right) \:.
\label{fmatrix1}
\end{eqnarray}
$\mathcal{D}U$ is the functional integration over the non-Abelian gauge link
variables $U_{n,\mu}$, $S_G$ is the discretized pure gauge action (we consider a
standard Wilson action); $u(B,q)_{i,\nu}$ are instead the
Abelian gauge links corresponding to the background e.m. field.  
The subscripts $i$ and $j$ refer to lattice
sites, $\hat\nu$ is a unit vector on the lattice and $\eta_{i,\nu}$ are the staggered
phases. Periodic (antiperiodic) boundary conditions (b.c.) must be taken, in the 
finite temperature theory, for gauge
(fermion) fields along the Euclidean time direction, while spatial 
periodic b.c. are chosen for all fields.

We shall consider
a constant and uniform magnetic field $\vec B = B \hat z$. The presence of periodic b.c.
in the $x$ and $y$ directions imposes a constraint
on the admissible values of $B$, which get quantized, as illustrated in the following subsection.
Symmetry under charge conjugation imposes that $Z(T,B)$ as well as other charge even observables, including
the chiral condensate, be even functions of $B$.

\subsection{Magnetic Field on a Torus}

\begin{figure}[h!]
\includegraphics*[width=0.47\textwidth]{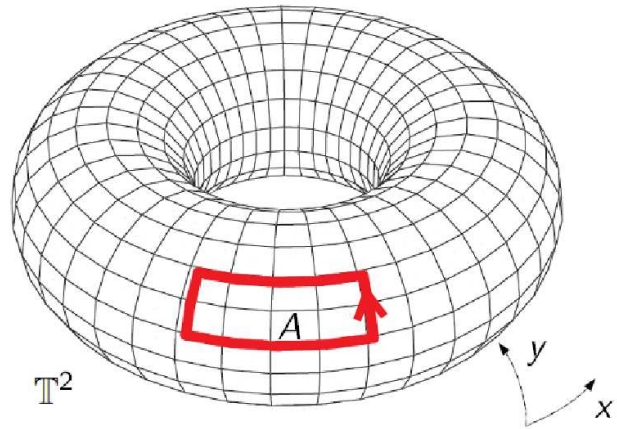}
\caption{Surface with periodic b.c. orthogonal to the direction of the magnetic field.
The phase factor taken by a particle moving around the plotted countour must be defined unambiguously
and this leads to magnetic flux quantization.
}
\label{torus}
\end{figure}

In presence of periodic b.c., the magnetic field in the $z$ 
direction goes through the surface of a torus in the $x-y$ directions, whose total
extent is $l_x l_y$. The circulation
of $A_\mu$ along any closed path, lying in the $x-y$ plane and enclosing an arbitrary region of area $A$ (see e.g. Fig.~\ref{torus}), is proportional, by Stokes' theorem, to 
the flux of $B$ through the enclosed surface 
\beq
\oint A_\mu d x_\mu = A B
\eeq
On the other hand, since we are on a torus, it is ambiguous to state which is the enclosed
surface: the complementary region of area $l_x l_y - A$ can be chosen as well,  
therefore one can equally state
\beq
\oint A_\mu d x_\mu = (A - l_x l_y) B \, .
\eeq
At the level of the gauge field the ambiguity is resolved by admitting discontinuities 
in $A_\mu$ somewhere on the torus, or alternatively by covering the torus with various
patches where different gauge choices are taken. In any case, one has to guarantee that 
the ambiguity is not visible by charged particles moving on the 
torus, and this is true only if the phase factor taken by the charged particle
moving along the closed path is defined unambiguously 
\beq
\exp \left( i q B A \right) = \exp \left( i q B (A - l_x l_y) \right) 
\eeq
i.e. if 
\beq
q B=2\pi b/l_x l_y
\label{defb}
\eeq
where $b$ is an integer. Notice that this line of argument is exactly the same that
applies on a sphere and that leads to Dirac quantization of the magnetic monopole charge.
The quantization rule depends on the 
electric charge of the particles feeling the presence of the magnetic field, in particular
it is set by the smallest charge unit, which in our case is brought by the $d$ quark,
$q_d = -|e|/3$,
hence
\beq
|e| B = 6 \pi b/l_x l_y 
= 6 \pi b a^{-2}/L_x L_y 
\label{Bquant}
\eeq
where $L_x$ and $L_y$ are the system sizes in lattice units and $a$ is the lattice spacing.

Further details about the definition of a magnetic field on a torus can be found in
Ref.~\cite{wiese}, where it is shown that translational invariance 
on the torus is explicitly broken by the presence of the magnetic field, due to the non-trivial phases
taken by particles winding around one of the directions of the torus (Wilson lines): only a discrete invariance is left, by shifts which are 
integer multiples of 
\beq
a_x = l_x/b \,\,\, ; \,\,\,\,\,\,\,\, a_y = l_y/b
\label{discrete_shifts}
\eeq
respectively in the $x$ and $y$ directions. Such invariance is reduced further
on a lattice, since only shifts, if any, which are multiples of both 
$a_x$ ($a_y$) and the lattice spacing $a$ leave the system invariant:
that may lead to additional discretization effects.

\subsection{Discretization details}

We have taken 
the following choice for the continuum e.m. gauge field:
\beq
A_y = B x
\:; \qquad\qquad
A_\mu = 0
\quad \mathrm{for} \quad \mu = x,z,t
\:.
\eeq
The corresponding $U(1)$ links on the lattice are:
\beq
u(B,q)_{n,y} = e^{i a^2 q B n_x}
\, ; \,\,\,
u(B,q)_{n,\mu} = 1
\: \mathrm{for} \: \mu = x,z,t\
\label{u1field}
\eeq
In order to guarantee the
smoothness of the background field across the boundary and the gauge invariance of
the fermion action, the $U(1)$ gauge fields must be modified at the boundary of the
$x$ direction:
\beq
u(B,q)_{n,x}|_{n_x = L_x} = e^{-i a^2 q L_x B n_y}
\label{boundary}
\:
\eeq
and the magnetic field must be quantized as specified in Eq.~(\ref{Bquant}).
That corresponds to taking the appropriate gauge 
invariant b.c. for fermion fields on the torus~\cite{wiese} 
(with the possible additional free phases
$\theta_x$ and $\theta_y$~\cite{wiese} set to zero).

We have considered a symmetric lattice, $L_x = L_y = L_z = N_t = 16$,
a bare quark mass $a m = 0.01335$ and an inverse gauge coupling $\beta = 5.30$.
According to scale estimates reported in Ref.~\cite{demusa}, that corresponds to 
a lattice spacing $a \simeq 0.3$ fm, a (Goldstone) pion mass $m_\pi \simeq 200$ MeV
and a temperature $T = (N_t a)^{-1} \simeq 40$ MeV, 
hence low enough that the system
can be considered to be effectively at zero temperature.

We have explored different values of $|e| B$ which, 
according to Eq.~(\ref{Bquant}),
can be changed only in units of $6 \pi a^{-2}/L_x L_y \simeq (180$~MeV)$^2$.
Notice however that the presence of an ultra-violet (UV) cutoff imposes
also an upper limit on the possible values of $B$ that can be explored 
on the lattice. To appreciate that, let us consider again the phase
factor picked up by a particle moving around a closed path
in the $x-y$ plane, and which contains all the relevant information about
the effect of the magnetic field on particle dynamics: there is a minimal such path on the lattice, corresponding
to a plaquette, around which the particle takes the phase factor
\beq
\exp (i q a^2 B) = \exp \left(  \frac{ i 2 \pi b}{L_x L_y} \right) \, .
\label{phasefactor}
\eeq
The phase factor above, and therefore all other phase factors associated to any
closed lattice path, cannot distinguish magnetic fields such that $ q a^2 B$ differs
by multiples of $2 \pi$. One can therefore define a sort of "first Brillouin zone"
for the magnetic field,
\beq
- \frac{\pi}{a^2} < q B < \frac{\pi}{a^2} 
\label{Bcutoff}
\eeq
i.e.
\beq
- \frac{L_x L_y}{2} < b < \frac{L_x L_y}{2} \, ,
\eeq
with all physical quantities being periodic in $q B$ $(b)$ with a period $2 \pi /a^2$ ($L_x L_y$); symmetry under $b \to -b$ further reduces the range
of interesting values of $b$. 
Even before reaching the limits reported in Eq.~(\ref{Bcutoff}), one expects the periodicity
to induce saturation effects, which may distort the true physical dependence of observables, like
the chiral condensate, on $B$. 
One should always worry about the possible presence of such saturation 
effects, when trying to extract information relevant to continuum physics.

\label{sec2B}

\subsection{Observables and simulation details}
\label{sec2C}

The quantity which is the subject of our investigation is the chiral
condensate. 
In presence of a non-zero $B$ 
we can define two different condensates
\beq
\label{observable}
\Sigma_{u/d}(B) &\equiv& \frac{\partial \log Z}{\partial m_{u/d}}
\vert_{m_{u/d} = m} \\ \nonumber
&=& \int \mathcal{D}U {\cal P} [m,U,B]
{\rm Tr} \left( M^{-1}[m,B,q_{u/d}]\right)
\eeq
where the functional integral measure is (see Eq.~(\ref{partfun1})):
\beq
\hspace{-3pt} {\cal P} [m,U,B] \hspace{-1pt}  \propto \hspace{-1pt} 
\det M^{1\over 4} [m,B,q_u] \hspace{-2pt}  
\det M^{1\over 4} [m,B,q_d]
e^{-S_{G}}
\label{measure}
\eeq

A quantity which is useful to discuss magnetic catalysis
is the relative increment of the quark condensate, which we define as:
\beq
r_{u/d}(B) \equiv \frac{\Sigma_{u/d}(B)}{\Sigma(0)} - 1 =
\frac{\Sigma_{u/d}(B) - \Sigma(0)}{\Sigma(0)} 
\, .
\label{defr}
\eeq
The advantage of $r(B)$ is that it is a dimensionless quantity and that most 
renormalizations appearing in the definition of $\Sigma$ cancels out
in Eq.~(\ref{defr}). Indeed, assuming that renormalizations 
have a negligible dependence on $B$, 
as should be the case as long as $B$ stays away from the scale of 
the UV cutoff, the mass dependent additive
renormalization of $\Sigma$ will cancel out in the numerator of 
Eq.~(\ref{defr}). A residual additive renormalization remains in the 
denominator, leading to an incorrect overall normalization of $r(B)$:
in the following we shall try to estimate the magnitude of such systematic 
error.

We shall define flavor averaged quantities as well:
\beq
\Sigma (B) = \frac{\Sigma_u (B) + \Sigma_d (B)}{2}
\label{avsigma}
\eeq
and
\beq
r(B) = \frac{\Sigma(B)}{\Sigma(0)} - 1 = \frac{r_u (B) + r_d (B)}{2}
\label{avr}
\eeq

Both $\Sigma_u(B)$ and $\Sigma_d(B)$ are, by charge conjugation symmetry, 
even functions of $B$.
Moreover, according to what discussed in Sec.~\ref{sec2B},
they are expected to be periodic
in $B$, with a period $2 \pi a^{-2}/q_d$ (or alternatively with a period $L_x L_y$ in terms of the
quantum number $b$ defined in Eq.~(\ref{defb})). One could expect the 
$u$ quark condensate to
have a periodicity shorter by a factor 2, since $|q_u| = 2 |q_d|$, however 
this is not exactly true because of the measure ${\cal P} [m,U,B]$ 
appearing in Eq.~(\ref{observable}), whose periodicity is set
by the quark with the lower charge.

In the limit $m \to 0$ the chiral condensate is an order parameter
for chiral symmetry breaking and is related by the Banks - Casher relation~\cite{banks}
to the density $\rho(\lambda)$
of eigenvalues of the Dirac operator, $D = M - m\ {\rm Id}$, around $\lambda = 0$: 
$\Sigma = \pi \rho(0)$. On the 
contrary, for $m \neq 0$, the chiral condensates defined
in Eq.~(\ref{observable}) are not related to the densities of zero 
eigenvalues of the respective Dirac operators, $\rho_{u/d} (0)$,
which instead can be obtained by taking the limit $m \to 0$ only for 
the trace term appearing in  Eq.~(\ref{observable}), i.e.
\beq
\rho_{u/d} (0) = \frac{1}{\pi} \lim_{m' \to 0} 
\int \mathcal{D}U {\cal P} [m,U]
{\rm Tr} \left( M^{-1}[m',q_{u/d}]\right)
\label{defrho}
\eeq
where the dependence on $B$ has been left implicit.
We shall consider also such quantities and the corresponding
relative increments 
\beq
\tilde r_{u/d} \equiv \rho_{u/d} (0,B)/\rho(0,B=0) - 1 \, .
\label{defrhorel}
\eeq

As discussed in the introduction, we are also interested in studying contributions to 
magnetic catalysis coming separately either from the change in the observable ${\rm Tr} \left( M^{-1}[m,B,q_{u/d}]\right)$ ("valence" contribution), or from
that in the measure ${\cal P} [m,U,B]$ ("dynamical" contribution).
For that reason we define also:
\beq
\Sigma^{val}_{u/d}(B) \equiv \int \mathcal{D}U {\cal P} [m,U,0]
{\rm Tr} \left( M^{-1}[m,B,q_{u/d}]\right)
\label{valence}
\eeq
and
\beq
\Sigma^{dyn}_{u/d}(B) \equiv \int \mathcal{D}U {\cal P} [m,U,B]
{\rm Tr} \left( M^{-1}[m,0,q_{u/d}]\right) \, .
\label{dynamical}
\eeq
In the first case we look at the spectrum of the fermion matrix which
includes the magnetic field explicitly, but is defined on 
non-Abelian configurations sampled at $B = 0$.
In the second case we look at the spectrum of the fermion matrix without an
explicit magnetic field, but defined on gauge configurations sampled in presence of the 
magnetic field. 
From $\Sigma^{val}_{u/d}(B)$ and $\Sigma^{dyn}_{u/d}(B)$ we can define the corresponding quantities,
$\Sigma^{val/dyn}, r_{u/d}^{val/dyn}, r^{val/dyn}$, analogously to what done in 
Eqs.~(\ref{defr}), (\ref{avsigma}) and (\ref{avr}).

On general grounds we may expect that, in the limit of small fields, $B$ acts as a perturbation
for both the measure term ${\cal P} [m,U,B]$ 
and the observable ${\rm Tr} \left( M^{-1}[m,B,q_{u/d}]\right)$
in Eq.~(\ref{observable}). Given that both functions
are even in $B$ and assuming they are also analytic (this may not be true in some limits, see
discussion below), so that the first non-trivial term in $B$ is quadratic,
one can write, configuration by configuration:
\beq
{\cal P} [m,U,B] =  {\cal P} [m,U,0] + C\ B^2 + O(B^4)
\label{smallp}
\eeq
and
\beq
\hspace{-6pt} {\rm Tr} \left( M^{-1}[B]\right) = {\rm Tr} \left( M^{-1}[0]\right) + C'\ B^2 + O(B^4) \, 
\label{smallt}
\eeq
where it is assumed implicitly that the two constants $C$ and $C'$ depend on the quark mass and 
on the chosen configuration.
Putting together the two expansions one obtains
\beq
\frac{\Sigma_{u/d}(B)}{\Sigma(0)} -1 = r^{val}_{u/d}(B)  + r^{dyn}_{u/d}(B)  + O(B^4) \, .
\label{additivity}
\eeq
Therefore, at least in the limit of small fields, the separation of magnetic catalysis
in a valence part and in a dynamical part is a well defined concept. 
As we will show in the following, that continues to be true, within a good approximation,
for a large range of fields explored in the present study. Notice that an approximate additivity
of  $r^{val}$ and $r^{dyn}$, like in Eq.~(\ref{additivity}), would be true also for 
different small field dependences, e.g. linear, in Eq.~(\ref{smallp}) and Eq.~(\ref{smallt});
hence the assumption above is stronger and also implies
that magnetic catalysis should be a quadratic effect in $B$, at least for small fields and if 
the partition function is analytic in $B = 0$.

We have made use of a Rational Hybrid Monte-Carlo algorithm to simulate rooted staggered fermions.
Typical statistics are of the order of 3k
thermalized molecular dynamics trajectories for each value of the magnetic field. 
The trace of the inverse of the fermion matrix, appearing in Eq.~(\ref{observable}),
has been computed, for each quark flavor, 
by means of a noisy estimator, extracting 10 random vectores for each configuration 
and for each value of the parameters.
Numerical simulations have been performed on the apeNEXT facilities in Rome.

\begin{table}
\begin{center}
\begin{tabular}{|c|c|c|c|c|c|}
\hline
$b$ &   $r_u(b)$ &  $r_d(b)$  & $r^{val}_u(b)$  & $r^{val}_d(b)$  &   $r^{dyn}_{u/d}(b)$ \\ \hline
1   &   0.0005(18) &  -0.001(2) & 0.0017(20)   &  0.0008(20) & -0.001(2)  \\
2   &   0.0077(19) &  0.0022(20) & 0.0070(19)   &  0.0027(18) & 0.0003(21)  \\
3   &   0.0202(16) &  0.0077(18) & 0.0151(20)   &  0.0037(19) & 0.0046(21)  \\
4   &   0.0356(22) &  0.0162(23) & 0.0266(19)   &  0.0052(19) & 0.0097(23)  \\
5   &   0.0567(18) &  0.0274(20) & 0.0407(19)   &  0.0121(19) & 0.0162(26)  \\
6   &   0.0760(19) &  0.0358(20) & 0.0579(20)   &  0.0165(18) & 0.0182(23)  \\
7   &   0.0996(16) &  0.0481(16) & 0.0759(20)   &  0.0217(19) & 0.0273(18)  \\
8   &   0.1246(17) &  0.0613(18) & 0.0949(19)   &  0.0281(19) & 0.0361(20)  \\
9   &   0.1474(16) &  0.0717(18) & 0.1144(19)   &  0.0352(18) & 0.0413(18)  \\
10  &   0.1736(17) &  0.0864(17) & 0.1340(19)   &  0.0412(18) & 0.0470(19)  \\
11  &   0.2005(18) &  0.1021(18) & 0.1554(19)   &  0.0503(19) & 0.0594(23)  \\
12  &   0.2258(16) &  0.1173(16) & 0.1765(19)   &  0.0584(19) & 0.0655(19)  \\
13  &   0.2501(17) &  0.1312(17) & 0.1983(20)   &  0.0676(18) & 0.0733(22)  \\
14  &   0.2737(18) &  0.1450(17) & 0.2192(20)   &  0.0762(20) & 0.0802(25)  \\
16  &   0.3227(19) &  0.1769(18) & 0.2568(20)   &  0.0957(19) & 0.0971(21)  \\
24  &   0.4636(23) &  0.2830(25) & 0.3809(21)   &  0.1777(19) & 0.1399(34)  \\
32  &   0.5462(22) &  0.3727(22) & 0.4472(24)   &  0.2594(21) & 0.1722(28)  \\
48  &   0.6485(22) &  0.5053(22) & 0.5308(23)   &  0.3816(21) & 0.2027(28)  \\
64  &   0.6855(23) &  0.5790(23) & 0.5652(24)   &  0.4460(21) & 0.2199(30)  \\
80  &   0.6545(22) &  0.6198(23) & 0.5317(23)   &  0.4924(22) & 0.2159(28)  \\
96  &   0.5726(21) &  0.6504(22) & 0.4480(22)   &  0.5297(22) & 0.2128(26)  \\
112 &   0.3868(19) &  0.6589(22) & 0.2603(21)   &  0.5549(23) & 0.1876(22)  \\
128 &   0.1333(18) &  0.6376(22) & 0.0000(19)   &  0.5642(22) & 0.1358(25)  \\
144 &   0.3828(21) &  0.6567(22) & 0.2583(21)   &  0.5558(22) & 0.1848(23)  \\ \hline
\end{tabular}
\end{center}
\caption{Relative increment of the $u$ and $d$ quark condensates for various magnetic field values. We report full data, as well as valence and dynamical contributions separately.}
\label{up_condensate}
\end{table}

\section{Numerical Results}
\label{results}

We report in Table~\ref{up_condensate} the relative increments of the 
$u$ and $d$ quark condensates respectively, including also measurements of the valence and of the dynamical
contribution. Notice that $r^{dyn}(B)$ is exactly the same, by definition, 
for $u$ and $d$ quarks, since
in this case the magnetic field affects only the fermion determinants.
One can also explicitly verify from the table
that $r^{val}_u(B/2) = r^{val}_d(B)$ within errors: this is expected since $|q_u| = 2 |q_d|$
and since in this case the sampling measure is independent of $B$.

\begin{figure}[h!]
\includegraphics*[width=0.47\textwidth]{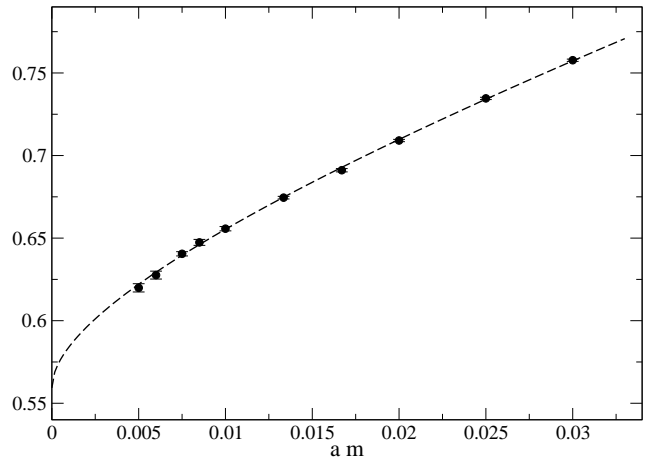}
\caption{Dependence of the chiral condensate on the bare quark mass at 
$B = 0$, together with a best fit curve according to Eq.~(\ref{massdep}).}
\label{cond_b0}
\end{figure}

As stressed in Section~\ref{sec2C}, all definitions of $r$ are affected 
by a systematic overall normalization factor, due to 
an additive, mass dependent renormalization present in the condensate computed 
at zero magnetic field, $\Sigma(0)$ (see 
Eq.~(\ref{defr})). In order to estimate the magnitude of such systematic
effect, we have performed simulations at $B = 0$ and different values of the 
quark mass, keeping the lattice size unchanged, in order to determine 
the dependence of $\Sigma(B=0,m)$ on $m$; results are reported
in Fig.~\ref{cond_b0}. The expected leading order dependence on $m$ 
in the chirally broken phase is the following 
(see e.g. the discussion in Ref.~\cite{ejiri}):
\beq
\Sigma(B = 0,m) = \Sigma(0,0) + c_{1/2} \sqrt{m}
+ c_1 m + O(m^3) 
\label{massdep}
\eeq
where the non-analytic square root term is expected from the presence
of Goldstone mode fluctuations~\cite{gold1,gold2,gold3}, while the leading
order, quadratically divergent contribution to the additive renormalization
affects the linear term in $m$.
A fit according to Eq.~(\ref{massdep}) gives
$\Sigma(B=0,0) = 0.565(6)$, $c_{1/2} = 0.61(9)$ and 
$c_1 = 2.9(4)$, with $\chi^2/{\rm d.o.f.} = 5.7/7$, from which
we infer that the linear term in $m$ accounts for about  6\% of the 
total signal measured at the quark mass explored in our investigation,
i.e. $am = 0.01335$. 

We conclude that our determinations
of $r$, $r^{val}$ and $r^{dyn}$ are distorted by a common and $B$ independent
overall normalization factor, which leads to a systematic effect of the order
of 10\% and does not affect issues such as the separation of 
magnetic catalysis into a valence and dynamical contribution,
as discussed later in this Section.

\subsection{Periodicity and saturation effects}

\begin{figure}[h!]
\includegraphics*[width=0.47\textwidth]{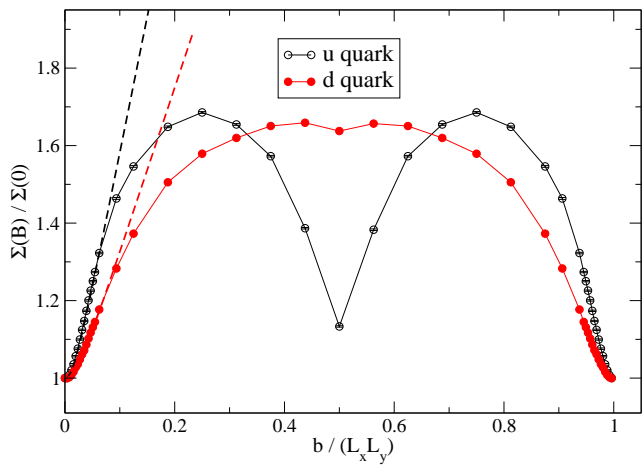}
\caption{Normalized $u$ and $d$ quark condensates as a function of the magnetic
field for the whole range of independent possible values of $B$. Data for 
$b/(L_x L_y) > 0.6$ have been obtained by enforcing the expected symmetry
of the chiral condensate under $b/(L_x L_y) \to 1 - b/(L_x L_y)$, while 
such symmetry has been verified for a couple of points, $b/(L_x L_y) = 0.4375$
and $b/(L_x L_y) = 0.5625$.
We also report two curves corresponding to best fits in the small field region, 
to better show the presence 
of saturation effects.}
\label{saturation}
\end{figure}

In Fig.~\ref{saturation} we report results for the normalized condensates $\Sigma_{u/d}(B)/\Sigma(B)$
(i.e. $1 + r_{u/d}(B)$)
over the whole range of possible independent values of $B$, i.e. for $b/(L_x L_y)$ ranging from 
0 to 1 (see discussion at the end of Sec.~\ref{sec2B}). Notice that 
data reported for 
$b/(L_x L_y) \geq 0.625$ are not the result of direct simulations, but have been obtained
by enforcing the expected invariances under $b \to b + L_x L_y$ (i.e. the above mentioned periodicity)
and under $b \to -b$, which put together mean invariance under $b/(L_x L_y) \to 1 - b/(L_x L_y)$.
However, such invariance has been verified explicitly for a couple of points, $b/(L_x L_y) = 0.4375$ and
$b/(L_x L_y)  = 0.5625$, for which independent simulations have been performed.

Saturation effects, which are present for large values of $B$, are clearly visible from 
Fig.~\ref{saturation}, where we have also reported for comparison the results of two fits
to the small $B$ region. In particular, we infer 
from the figure that one should keep $b/(L_x L_y)$ well below 0.1 in order that such effects
stay negligible: that means $|e|B$ below $1/a^2 \sim (700\ {\rm MeV})^2$ in our case.
In the following only data obtained for $b \leq 16$ will be considered
as reasonably free of saturation effects.

Notice that the $u$ quark condensate shows an approximate periodicity in $B$
which is halved with respect to the $d$ quark. That comes from the fact that 
$|q_u| = 2 |q_d|$
and is only approximate since instead the measure term has the usual periodicity. For instance
$r_u(b)$ has a minimum but is not exactly zero 
at $b/(L_x L_y) = 1/2$, where the effective magnetic field felt by $u$ quarks is zero:
 there is a residual catalysis induced by dynamical $d$ quarks, 
which instead at $b/(L_x L_y) = 1/2$ feel the maximum
possible magnetic field; for the same reason $r_d(b)$ does not reach
its maximum at $b/(L_x L_y) = 1/2$.
Such effects, which are absent for the purely valence contribution as can
be checked from Table~\ref{up_condensate}, are a first example of the dynamical contribution to magnetic
catalysis, which we discuss in detail in the next subsection.

\subsection{Dynamical and Valence contributions to magnetic catalysis}
\label{sec3B}

\begin{figure}[h!]
\includegraphics*[width=0.47\textwidth]{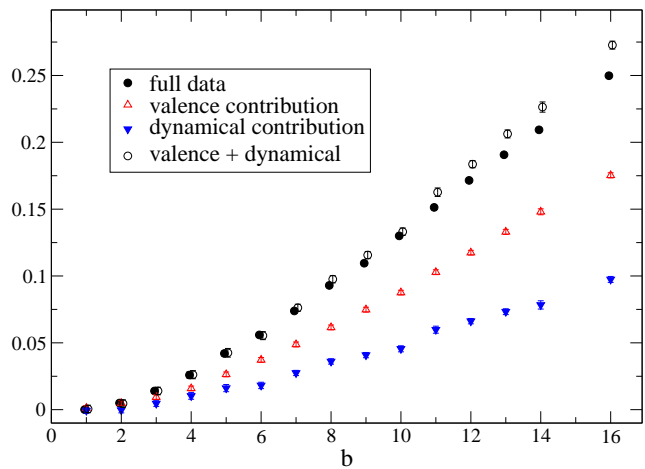}
\caption{Relative increment of the average of the $u$ and $d$ quark condensates 
as a function of the magnetic field. We report separately $r(B), r^{val}(B), r^{dyn}(B)$
and $r^{val}(B) + r^{dyn}(B)$.}
\label{comparison}
\end{figure}

In Fig.~\ref{comparison} we report the functions $r(B), r^{val}(B), r^{dyn}(B)$
(see Eqs.~(\ref{valence}) and (\ref{dynamical})) 
as well as the sum $r^{val}(B) + r^{dyn}(B)$, in order to appreciate the amount
of magnetic catalysis caused by the modified distribution
of the non-Abelian gauge fields, induced by the coupling of dynamical quarks
to the magnetic field (dynamical contribution).
We have limited our analysis to $b \leq 16$, for which  
saturation effects do not play a significant role.

The first thing that we notice is that the dynamical and valence contributions
are roughly additive, in the sense that their sum gives back to full signal,
in the range of fields shown in the figure. The additivity, which is expected 
in the limit of small fields (see discussion in Sec.~\ref{sec2C}), is verified
within errors for $b \leq 8$ ($|e| B \leq  (500\ {\rm MeV})^2$), while small deviations
appear beyond. Notice that this threshold coincides with that above which 
a purely quadratic fit for $r(B)$ does not work (see next subsection) and 
quartic terms in $B$ become important, in agreement
with the argument given in  Sec.~\ref{sec2C}.

Once clarified that it is sensible, in the explored range of fields, to divide 
magnetic catalysis into a valence and a dynamical contribution,
from Fig.~\ref{comparison} we learn that the dynamical one is roughly 
40\% of the total signal, at least for the discretization 
and quark mass spectrum adopted in our investigation.
That means that numerical studies in which the magnetic
field is not included in the sampling distribution (quenched or partially quenched)
may miss a substantial part of magnetic catalysis, in a measure larger
than other systematic effects due to quenching, which are typically of the order of 20\%.

\begin{figure}[h!]
\includegraphics*[width=0.47\textwidth]{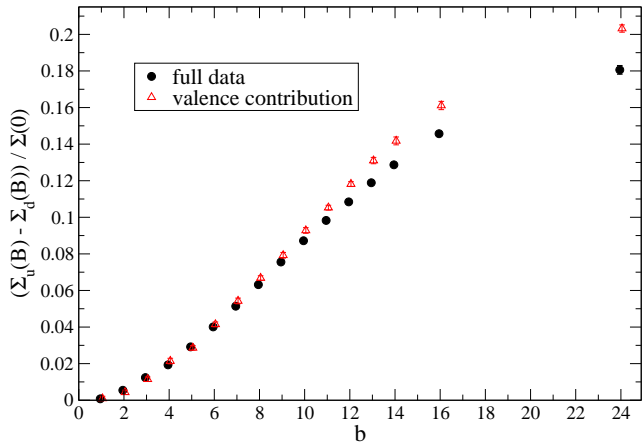}
\caption{Difference of the $u$ and $d$ quark condensates, normalized by the 
zero field condensate, as a function of the magnetic field and computed respectively
on configurations sampled by taking or not taking into account the magnetic field
in the fermion determinant.}
\label{diffud}
\end{figure}

In Fig.~\ref{diffud} we have also plotted results for the difference between the $u$ and the $d$ condensates,
which increases as a function of $B$, indicating an increasing 
breaking of flavor symmetry. The fact that the dynamical contribution is equal
for the two quarks and the approximate additivity discussed above implies that such difference
should be roughly unchanged if we consider just the valence contribution: that can be verified
again from Fig.~\ref{diffud}.

\subsection{Comparison with $\chi$PT and model predictions.}

One of the purposes of our investigation is to compare
our results with various analytic studies based 
on low energy or model approximations of QCD. Most of those studies
make reference to the average quark condensate and not to the 
$u$ or $d$ condensates separately.
Among the various existing  predictions, 
one of the first was based on the analysis of the 
Nambu - Jona-Lasino model~\cite{Klevansky:1989vi} and predicted a quadratic increase
of the condensate as a function of the magnetic field, i.e.
$r(B) \propto B^2$.

A first prediction based on chiral perturbation theory has been proposed
in Ref.~\cite{Shushpanov:1997sf}
\begin{equation}
\label{smilga} 
r(B) = 
\frac{\log(2) \, |e| B}{16 \pi^2 F_{\pi}^2}.
\end{equation}
and is valid only in the chiral limit, i.e. $m_\pi = 0$, and
for $|e|B \ll \Lambda_{\rm QCD}^2$; in the limit 
of strong fields, instead, the authors of Ref.~\cite{Shushpanov:1997sf}
have predicted a power law behavior
$r(B) \propto |B|^{3/2}$~\cite{Shushpanov:1997sf}.
Corrections to Eq.~(\ref{smilga}), based on a two-loop computations, 
have been given in Ref.~\cite{Agasian:1999sx}. 

The authors of Ref.~\cite{Cohen:2007bt} have gone beyond the limitation
$m_\pi = 0$, presenting a $\chi$PT computation which is valid
for generic values of $m_\pi^2/(|e| B)$, even if still for 
$|e|B \ll \Lambda_{\rm QCD}^2$. The prediction in this case is:
\beq 
\label{cohen}
r(B) =   \frac{\log (2)\, eB}{16 \pi^2 F_{\pi}^2} I_H
\left( \frac{m_{\pi}^2}{ |e| B} \right )
\eeq
where
\beq 
\label{Ih}
I_H (y) &=& \frac{1}{\log 2}\left( \log(2 \pi) +y\log \left(\frac{y}{2}\right) \right.
\nonumber \\
 && \ \ \ \ \ \ \ \ \ \ -  \left. y
- 2 \log \Gamma\left(\frac{1 + y}{2}\right)\right) \, .
\eeq
Notice that $I_H(y)  \to 1$ as $y \to 0$, i.e. in the chiral limit,
in agreement with Eq.~(\ref{smilga}).

Recently various predictions have been proposed,
based on the holographic AdS/CFT 
correspondence~\cite{Johnson:2008vna,Bergman:2008qv,
Zayakin:2008cy,Evans:2010xs}: the increase in chiral symmetry breaking is 
confirmed in all cases, with a dependence of the chiral condensate on $B$ 
which ranges from quadratic~\cite{Zayakin:2008cy} to a power law,
e.g. $r(B) \propto |B|^{3/2}$~\cite{Evans:2010xs}.

Existing lattice determinations have reported
a linear behaviour for SU(2) pure gauge theory~\cite{itep1} and 
a power law behavior $r(B) \propto B^\nu$ (with $\nu \sim 1.6$) for the 
SU(3) pure gauge theory~\cite{itep2}.

Regarding the small field behavior, 
one can state on general grounds that, since by charge conjugation symmetry
the chiral condensate must be an even function of $B$, if the theory is 
analytic at $B = 0$ then the chiral condensate can be written as a Taylor 
in expansion in $B^2$, hence for small enough fields the dependence must 
be quadratic. 

That is indeed in agreement with many model predictions
and is not true only in some particular cases: for instance
the prediction from $\chi$PT in Eq.~(\ref{smilga})~\cite{Shushpanov:1997sf}
is linear, since the analyticity requirement is violated in 
this case due to the fact that the pion mass is set to zero.
 Let us consider instead the prediction from 
Ref.~\cite{Cohen:2007bt}, which is valid for generic values of  
$m_\pi^2/(|e| B)$. While for $m_\pi^2/(|e| B) \to 0$ Eq.~(\ref{cohen})
gives back a linear behavior as in Eq.~(\ref{smilga}), in the opposite limit
$|e| B \ll m_\pi^2$,
i.e. ($\frac{1}{y}<<1$), we find, by expanding Eq.~(\ref{Ih})
in powers of
powers of $\frac{1}{y}$, that
\beq
\label{low_H}
r(B) \simeq \frac{(|e|B)^2}{96 \pi^2 F_\pi^2 m_\pi^2} 
\eeq
in agreement with the general expectation. Such behavior, quadratic
for small fields and linear for larger fields, has been found also in a recent
study based on the linear sigma model~\cite{mizher}.

\begin{figure}[h!]
\includegraphics*[width=0.47\textwidth]{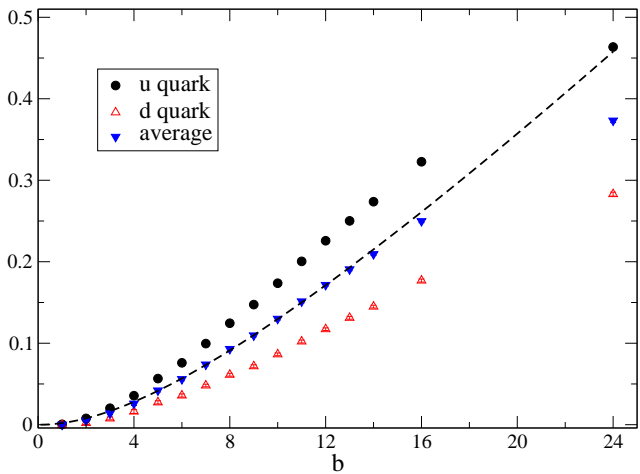}
\caption{Relative increment of the quark condensate as a function of the 
magnetic field. We report separately data for the $u$ and $d$ quarks as well as for
the average of the two, together with our best fit reported in the 6th line of 
Table~\ref{Fit_Cohen}.}
\label{fulldata}
\end{figure}

In Fig.~\ref{fulldata} we report the relative increment of the $u$ and 
$d$ condensates and of their average in a restricted
region for which we expect that saturation
effects are not important. We remind that in our case 
the variable representing the magnetic field is the dimensionless parameter
$b$, and the conversion to physical units is given by Eq.~(\ref{Bquant}),
which for our lattice size reads $|e| B = (6 \pi /256)\ b/a^2$,
with $a \simeq 0.3$ fm, i.e. $|e| B \simeq b\ (180\ {\rm MeV})^2$.

It is apparent by eye that Eq.~(\ref{smilga}) badly fits with our data, indeed a linear
fit gives unacceptable values for the $\chi^2/{\rm d.o.f.}$ test
(e.g. $\chi^2/{\rm d.o.f.} \simeq 190/6$ for $1 \leq b \leq  7$).
This is expected, since in our case $m_\pi \neq 0$.
However, data at larger values of $B$ show an approximate linear behavior and
if we discard the smallest values of $b$ a function $r(b) = a_0 + a_1\ b$ reasonably
fits our data: for instance, a fit including $b_{\rm min} = 6$ and $b_{\rm max} = 16$
gives $a_0 = -0.063(2)$, $a_1 = 0.0195(2)$ and $\chi^2/{\rm d.o.f.} \simeq 0.68$.

Next we have tried to check if a quadratic behavior 
$r(B) = (|e| B / \Lambda_B^2)^2$ fits better, at least for small enough fields. Results
are reported in Table~\ref{Fit_quadr}. For $b < 8$, i.e.
$|e| B < (500\ {\rm MeV})^2$, the quadratic fit looks good
and stable, with $\Lambda_B \sim 900$ MeV. That is in agreement
with the general expectation for the  case of small fields discussed above
(even if $500$ MeV is not small with respect to $\Lambda_{\rm QCD}$). 
We notice that, in agreement with
the argument given in Sec.~\ref{sec2C}, the range of validity of the quadratic
fields roughly coincides with the range in which the dynamical and the valence
contribution are additive (see Sec.~\ref{sec3B}).

\begin{table}
\begin{center}
\begin{tabular}{|c|c|c|c|}
\hline 
  $b_{\rm min}$    & $b_{\rm max}$ &  $\chi^2/\textnormal{d.o.f} $ &    $\Lambda_B \;(MeV)$    \\ [0.5ex]
\hline  
  1       &   4  &   0.45                      &      903(11)      \\  
  1       &   5  &   0.52                      &      892(6)       \\
  1       &   6  &   0.67                      &      899(5)       \\
  1       &   7  &   0.98                      &      907(4)       \\
  1       &   8  &   1.7                       &      914(4)       \\
  1       &   9  &   4.6                       &      923(5)       \\
  1       &  10  &   8.4                       &      933(5)       \\
  1       &  11  &   12                        &      940(5)       \\
  1       &  12  &   21                        &      950(6)       \\
  1       &  13  &   32                        &      960(6)       \\
  1       &  14  &   48                        &      969(7)       \\
  1       &  16  &   86                        &      983(8)       \\
\hline
\end{tabular}
\end{center}
\caption{Results from a quadratic fit $r(B) = (|e| B / \Lambda_B^2)^2$
to our data. $b_{\rm min}$ and 
$b_{\rm max}$ indicate respectively 
the minimum ad maximum values of the magnetic fields
included in the fit.} 
\label{Fit_quadr}
\end{table}

\begin{table}[t!]
\begin{center}
\begin{tabular}{|c|c|c|c|c|}
\hline
 $b_{\rm min}$    & $b_{\rm max}$ &  $\chi^2/\textnormal{ndf} $ & $m_\pi \;(MeV)$  & $F_\pi \;(MeV)$  \\ [0.5ex]
\hline  
 1       &   8  &   0.63                      &      457(59)     &   54.7(5.6)  \\
 1       &   9  &   0.93                      &      392(39)     &   61.7(4.2)  \\
 1       &  10  &   0.89                      &      374(27)     &   63.8(2.9)  \\
 1       &  11  &   0.80                      &      369(20)     &   64.3(2.1)  \\
 1       &  12  &   0.79                      &      359(15)     &   65.6(1.6)  \\
 1       &  13  &   0.97                      &      344(14)     &   67.2(1.4)  \\
 1       &  14  &   1.40                      &      328(14)     &   69.1(1.4)  \\
 1       &  16  &   1.92                      &      310(13)     &   71.1(1.3)  \\
 1       &  24  &   17.4                      &      227(22)     &   80.8(2.3)  \\
 4       &  14  &   1.00                      &      320(12)     &   69.9(1.2)  \\
 6       &  14  &   0.97                      &      313(13)     &   70.6(1.3)  \\
 8       &  14  &   0.83                      &      298(15)     &   72.2(1.5)  \\
\hline
\end{tabular}
\end{center}
\caption{Results from a fit of our data to Eq.~(\ref{cohen}). The notation
for $b_{\rm min}$ and 
$b_{\rm max}$ is as in Table~\ref{Fit_quadr}.}
\label{Fit_Cohen}
\end{table}

\begin{table}[t!]
\begin{center}
\begin{tabular}{|c|c|c|c|c|}
\hline 
  $b_{\rm min}$    & $b_{\rm max}$ &  $\chi^2/\textnormal{ndf} $ & $\Lambda_B \;(MeV)$  & $\nu$  \\ [0.5ex]
\hline
  1       &   4  &   0.23                      &      784(54)     &   2.33(19) \\
  1       &   5  &   0.19                      &      817(25)     &   2.23(9) \\
  1       &   6  &   0.84                      &      900(42)     &   2.00(12) \\
  1       &   7  &   0.92                      &      940(31)     &   1.90(8)  \\
  1       &   8  &   0.97                      &      965(24)     &   1.85(6)  \\
  1       &   9  &   1.7                       &     1006(26)     &   1.75(6)  \\
  1       &  10  &   1.9                       &     1031(21)     &   1.70(5)  \\
  1       &  11  &   1.9                       &     1045(18)     &   1.67(4)  \\
  1       &  12  &   2.3                       &     1064(15)     &   1.63(4)  \\
  1       &  13  &   3.1                       &     1083(15)     &   1.59(4)  \\
  1       &  14  &   4.3                       &     1104(16)     &   1.55(4)  \\
  1       &  16  &   6.4                       &     1131(16)     &   1.50(4)  \\
  1       &  24  &   35.8                      &     1254(30)     &   1.29(5)  \\
\hline
\end{tabular}
\end{center}
\caption{Results from a fit $r(B) = (|e| B / \Lambda_B^2)^\nu$.
The notation
for $b_{\rm min}$ and 
$b_{\rm max}$ is as in Table~\ref{Fit_quadr}.
}
\label{Fit_powerlaw}
\end{table}

We have then tried to fit our data with the prediction
of Ref.~\cite{Cohen:2007bt}, as reported in Eq.~(\ref{cohen}).
We have obtained reasonable fits only if both $m_\pi$ and
$F_\pi$ are treated as independent free parameters, results 
are reported in Table~\ref{Fit_Cohen}.
Data are well described by the prediction in Eq.~(\ref{cohen}) 
over a wide range of values of $|e| B$, including $b = 14$ 
(i.e. $|e| B \sim$ 700 MeV),
even if the fitted values
of $F_\pi$ and $m_\pi$ are not very stable as the range is modified.

Typical values of the fit parameters are 
$m_\pi \sim 300-400$~MeV and $F_\pi \sim 60-70$ MeV. 
The fitted pion mass is somewhat larger than the value
obtained, with the same discretization settings, 
by measuring meson correlators, i.e. $m_\pi \sim 200$ MeV~\cite{demusa},
however that is quite reasonable if we take into account the many 
discretization systematic effects which affect our simulations.
The most relevant comes from the explicit 
flavour symmetry breaking induced by the staggered
discretization: the three pions are
not degenerate in mass and what is determined by
meson correlator measurements is just the lowest pion mass:
it is therefore likely that the $\chi$PT prediction 
still holds, but with an heavier effective pion mass.

Regarding $F_\pi$, the fitted values are about
$\sim$ 20-30\% lower than the expected physical value,
$F_\pi \simeq 93$ MeV. We notice that $F_\pi$ enters
Eq.~(\ref{cohen}) only in the prefactor, 
hence its value is surely affected by the systematic
uncertainty in the overall normalization factor
for $r(B)$, which is of the order of 10\%.
There are also corrections expected from the fact that
we are not at zero temperature: the
authors of Ref.~\cite{Agasian:2001ym} predict
$F_\pi^2(T) \simeq F_\pi^2 - {T^2}/{6}$, however
in our case $T \simeq 40$ MeV and that can account 
for at most a 2\% discrepancy from the physical value.
The non-physical large value of $m_\pi$ can also affect 
$F_\pi$, but $\chi$PT would predict 
an increased value of $F_\pi$~\cite{chipt}.

There are however many other possible sources of systematic 
uncertainties, including the fact that the $\chi$PT prediction
of Ref.~\cite{Cohen:2007bt} has been obtained in the low energy
limit $|e| B \ll \Lambda_{\rm QCD}^2$, a condition which is violated 
in our explored range of fields.
We have verified that other two-parameter functions, which allow
to fix independently the curvature at $B = 0$ and the asymptotic linear behavior for
larger fields (like Eq.~(\ref{cohen}) when $F_\pi$ and $m_\pi$ are treated as
independent parameters) work equally well. For instance the 
function
\beq
r(b) = c_0 b\  {\rm atan} ( c_1 b)
\label{atan}
\eeq
fits well in the whole range of explored fields, $b_{\rm min} = 1$ and
$b_{\rm max} = 16$, with $c_0 = 0.0136(2)$, $c_1 = 0.140(5)$ and  
$\chi^2/{\rm d.o.f.} \simeq 0.77$.

To compare with the analysis performed in Ref.~\cite{itep2}, 
we have also investigated if a power law behaviour $r(B) = (|e| B / \Lambda_B^2)^\nu$ 
can fit our data. Results are reported in Table~\ref{Fit_powerlaw}. 
Reasonable fits are obtained only for a range of fields including $b  = 8$. That 
coincides more or less with the range for which also the quadratic fit works well, and indeed
values obtained for $\nu$ in this range are roughly compatible with $\nu = 2$.

We are not able to say much about the strong field regime, $|e| B \gg \Lambda_{\rm QCD}^2$,
since saturation effects make such regime unaccessible to our present investigation:
much smaller lattice spacings should be used to that aim.

\subsection{Density of zero modes} 

Let us finally discuss our results for the densities of zero modes of the 
Dirac operator, $\rho_u(0)$ and $\rho_d(0)$, obtained for the $u$ and $d$ 
quarks respectively. We have determined such densities following
Eq.~(\ref{defrho}), i.e. determining on our ensemble of configurations,
sampled with a dynamical quark mass $a m = 0.01335$,
the average of ${\rm Tr} (M^{-1}(a m'))$ for various different values
of $m'$ and then extrapolating to $a m' = 0$. In particular we have chosen
$a m' = 0.007, 0.01335, 0.02$ and $0.03$. 
A quadratic extrapolation in $m'$ works
well in all cases.

\begin{figure}[h!]
\includegraphics*[width=0.47\textwidth]{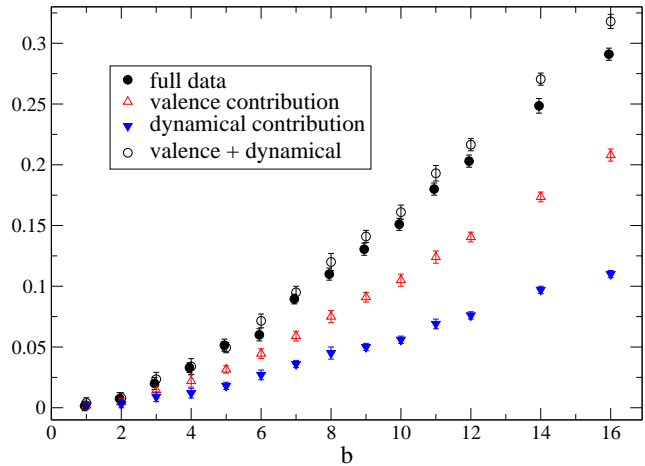}
\caption{Relative increment of the average of the $\rho_u(0)$ and $\rho_d(0)$  
as a function of the magnetic field. We report separately $\tilde r(B), \tilde r^{val}(B), \tilde r^{dyn}(B)$
and $\tilde r^{val}(B) + \tilde r^{dyn}(B)$.}
\label{comparison_rho}
\end{figure}

In Table~\ref{tab_zeromodes} we report data
for the relative increments $\tilde r_u$ and $\tilde r_d$ of $\rho_u(0)$ and $\rho_d(0)$ as a function
of $B$, also for the cases
in which only the valence or the dynamical contributions are taken into account;
the average quantities are plotted in Fig.~\ref{comparison_rho}.
We notice, comparing Table~\ref{tab_zeromodes} with Table~\ref{up_condensate}, that
the relative increment of the zero mode density, $\tilde r$, is generally larger
than the corresponding increment in the condensate, $r$. Moreover Fig.~\ref{comparison_rho}
shows that, similarly to what happens for the condensate, it is possible
to make a separation of the total increment into a valence and a dynamical
contribution, which sum approximately to the total increment. Also in this case
the change in the distribution of the non-Abelian gauge fields (dynamical contribution)
accounts for about 30-40\% of the total increase.

\begin{table}
\begin{center}
\begin{tabular}{|c|c|c|c|c|c|}
\hline
$b$ &   $\tilde r_u(b)$ &  $\tilde r_d(b)$  & $\tilde r^{val}_u(b)$  & $\tilde r^{val}_d(b)$  &   $\tilde r^{dyn}_{u/d}(b)$ \\ \hline
1   &   0.003(4) &  0.000(4) & 0.002(3)   &  0.002(3) & 0.002(3)  \\
2   &   0.013(4) &  0.002(5) & 0.007(3)   &  0.003(3) & 0.003(3)  \\
3   &   0.026(4) &  0.014(5) & 0.022(4)   &  0.007(4) & 0.009(4)  \\
4   &   0.042(3) &  0.024(4) & 0.034(4)   &  0.010(5) & 0.012(4)  \\
5   &   0.069(3) &  0.034(5) & 0.047(3)   &  0.016(3) & 0.018(3)  \\
6   &   0.080(3) &  0.040(7) & 0.068(3)   &  0.021(4) & 0.027(4)  \\
7   &   0.117(3) &  0.062(4) & 0.087(3)   &  0.031(4) & 0.036(3)  \\
8   &   0.144(5) &  0.076(5) & 0.113(6)   &  0.037(5) & 0.045(5)  \\
9   &   0.173(5) &  0.088(4) & 0.138(5)   &  0.044(4) & 0.050(3)  \\
10  &   0.201(3) &  0.101(7) & 0.158(5)   &  0.052(5) & 0.056(3)  \\
11  &   0.236(5) &  0.124(4) & 0.185(5)   &  0.063(5) & 0.069(4)  \\
12  &   0.264(4) &  0.142(5) & 0.207(4)   &  0.074(5) & 0.076(3)  \\
14  &   0.319(5) &  0.178(8) & 0.256(4)   &  0.091(4) & 0.097(3)  \\
16  &   0.372(5) &  0.211(5) & 0.302(6)   &  0.115(4) & 0.110(3)  \\
\hline
\end{tabular}
\end{center}
\caption{Relative increment of the density of zero modes of the Dirac operator for the $u$ and $d$ quarks and 
for various magnetic field values. We report full data, as well as valence and dynamical contributions separately.}
\label{tab_zeromodes}
\end{table}

\section{Conclusions}
\label{conclusions}

In this study we have approached the issue of magnetic catalysis by numerical lattice
simulation of $N_f = 2$ QCD. We have adopted a rooted standard staggered 
discretization of the fermion action and a plaquette pure gauge action
on a symmetric $16^4$ lattice, corresponding to a lattice spacing $a \simeq 0.3$ fm,
a (Goldstone) pion mass $m_\pi \simeq 200$ MeV and a temperature well below
the deconfinement/chiral restoring one ($T \sim 40$ MeV). We have studied the 
breaking of chiral symmetry as a function of a constant and uniform
magnetic field directed along the $\hat z$ axis. Explored magnetic fields, allowed
by the toroidal geometry and for which distortion (saturation) effects due to 
the lattice discretization are not significant, range from $|e| B \sim (180\ {\rm MeV})^2$
to $|e| B \sim (700\ {\rm MeV})^2$.

We have shown that, in the range of explored fields, it is possible to divide magnetic 
catalysis into a contribution coming from the modified distribution of non-Abelian
gauge fields, induced by dynamical quark loop effects, that we have called dynamical
contribution, and a valence contribution, determined by measuring the condensate on
gauge configurations sampled with the unmodified distribution. 
The first term, which is missed by quenched or partially 
quenched studies, accounts for about 40\% of the total increase in the quark condensate.
Results obtained for the density of zero modes looks quite similar.

Regarding the dependence of the condensate on the magnetic field, we have shown that
a quadratic behavior, which is expected in the limit of small magnetic fields, describes 
well our data for $|e| B$ up to $\sim$ (500 MeV)$^2$. The $\chi$PT prediction of 
Ref.~\cite{Cohen:2007bt} fits data over a wider range, but only if the pion
decay constant is treated as an independent free parameter.

Our investigation can be improved in several respects. Lattice artifacts may affect our 
results in various different ways, ranging from the presence 
of a distorted meson spectrum in the adopted
rooted staggered fermion formulation, to residual renormalization effects and possible
residual saturation effects in the explored range of fields.
An improved lattice
formulation and a finer spacing $a$ would allow to check for such
artifacts, to test the correct scaling to the continuum limit of our results 
and to explore larger values of the magnetic field. A larger spatial 
volume would instead allow for a finer quantization of $|e| B$ and a better investigation
of the small field region. It would be also interesting to explore different
choices of the quark mass spectrum, in order to see how the separation of magnetic
catalysis into a dynamical and a valence contribution depends on the dynamical quark masses.
We plan to address such issues in future studies.

\begin{acknowledgements}
We thank P.~Cea, M.~Chernodub, T.~Cohen, L.~Cosmai, V.~Miransky, M.~Ruggieri and F.~Sanfilippo
for interesting discussions. We are grateful to 
G.~Endrodi and S.~Mukherjee for interesting discussions and correspondence regarding renormalization effects.
\end{acknowledgements}

\end{document}